\documentclass[prd,twocolumn,nofootinbib,aps,tightenlines,preprintnumbers,notitlepage,longbibliography,superscriptaddress]{revtex4-1}
\usepackage{graphicx}
\usepackage{amsmath}
\usepackage{amsfonts}
\usepackage[colorlinks=true,citecolor=blue,urlcolor=cyan,linkcolor=black]{hyperref}

\newcommand{\nhat}{\hat{ \mathbf{n}}}

\usepackage{graphicx}
\usepackage{dcolumn}
\usepackage{bm}
\usepackage[usenames, dvipsnames]{color}
\usepackage[normalem]{ulem}
\usepackage{xcolor}

\graphicspath{ {figures/} }

\usepackage{epstopdf}
\newcommand{\be}{\begin{eqnarray}}
\newcommand{\non}{\nonumber \\}
\newcommand{\ee}{\end{eqnarray}}

\usepackage[export]{adjustbox}

\def\nhat{\hat{\mathbf{n}}}
\def\rhat{\hat{\mathbf{r}}}

\usepackage{suffix}
\usepackage{mathtools}

\DeclarePairedDelimiterX\MeijerM[3]{\lparen}{\rparen}%
{\begin{smallmatrix}#1 \\ #2\end{smallmatrix}\delimsize\vert\,#3}

\newcommand\MeijerG[8][]{%
  G^{\,#2,#3}_{#4,#5}\MeijerM[#1]{#6}{#7}{#8}}

\WithSuffix\newcommand\MeijerG*[7]{%
  G^{\,#1,#2}_{#3,#4}\MeijerM*{#5}{#6}{#7}}

\newcommand{\dd}{{\rm d}}
\newcommand{\bb}{{\rm b}}

\begin{document}

\title{Optimal filters for the moving lens effect}

\newcommand{\imperial}{Department of Physics, Imperial College London, Blackett Laboratory, Prince Consort Road, London SW7 2AZ, UK}

\newcommand{\perimeter}{Perimeter Institute for Theoretical Physics, 31 Caroline St N, Waterloo, ON N2L 2Y5, Canada}

\newcommand{\york}{Department of Physics and Astronomy, York University, Toronto, ON M3J 1P3, Canada}

\newcommand{\CITA}{Canadian Institute for Theoretical Astrophysics,
University of Toronto, Toronto, ON M5H 3H8 Canada}

\newcommand{\smu}{Department of Physics,
Southern Methodist University, 3215 Daniel Ave, Dallas, TX 75275, U.S.A.}

\author{Selim~C.~Hotinli}
\affiliation{\imperial}

\author{Matthew~C.~Johnson}
\affiliation{\perimeter}
\affiliation{\york}

\author{Joel~Meyers}
\affiliation{\smu}

\date{\today}

\begin{abstract}

We assess the prospects for detecting the moving lens effect using cosmological surveys. The bulk motion of cosmological structure induces a small-scale dipolar temperature anisotropy of the cosmic microwave radiation (CMB), centered around halos and oriented along the transverse velocity field. We introduce a set of optimal filters for this signal, and forecast that a high significance detection can be made with upcoming experiments. We discuss the prospects for reconstructing the bulk transverse velocity field on large scales using matched filters, finding good agreement with previous work using quadratic estimators. 

\end{abstract}

\maketitle

\section{Introduction}

Observations of the cosmic microwave background (CMB) with the upcoming Simons Observatory (SO)~\citep{Ade:2018sbj} and CMB-S4~\citep{Abazajian:2016yjj} experiments, along with galaxy surveys such as LSST (Vera~C.~Rubin Observatory)~\citep{2009arXiv0912.0201L}, will open new windows of opportunity for cosmological inference. In particular, there is evidence that the measurement of small-scale secondary anisotropies that are imprinted on the CMB by cosmological structures between our telescopes and the surface of last scattering will provide strong constrains on a multitude of cosmological signatures (see e.g.~\citep{Schmittfull:2017ffw,Deutsch:2018umo,Munchmeyer:2018eey,Cayuso:2019hen,Pan:2019dax,Ballardini:2019wxj,Hotinli:2019wdp}). The statistics of these secondaries and their cross-correlations with large-scale structure (LSS) surveys carry information about cosmological fluctuations on large scales. Utilizing this information will be instrumental in future tests of the standard $\Lambda$CDM paradigm. These secondary effects include weak gravitational lensing by large-scale structure; the integrated Sachs-Wolfe (ISW) and Rees-Sciama effects, which describe the redshifting of CMB photons due to evolving gravitational potentials along the line of sight; and the Sunyaev-Zel'dovich (SZ) effect where CMB photons scatter with free electrons in galaxy clusters and the intergalactic medium. In this work, we study the moving lens effect: temperature anisotropies in the CMB due to the peculiar velocity of cosmological structure transverse to the line of sight~\citep{1983Natur.302..315B,1986Natur.324..349G,Aghanim:1998ux,Cooray:2002dia,Lewis:2006fu}. It has recently been shown (see~e.g.~\citep{Hotinli:2018yyc,Yasini:2018rrl}) that this effect can in principle be detected at high-significance for the first time with upcoming surveys. 

A major goal of the scientific program of measuring secondaries is constraining fundamental physics. Large-scale cosmological perturbations leave unique imprints on the small-scale intensity and polarization anisotropies of the CMB. The study of these  statistical anisotropies provide new information about the largest scales in the Universe. Large-scale observables are in turn valuable for cosmological inference as they are often protected from local and non-linear late time effects under the equivalence principle.\footnote{Equivalence principle dictates that local interactions produce density fluctuations that scale with the Fourier wavenumber $k$ like the Laplacian (or the time derivative) of the gravitational potential $\nabla^2\Phi$ (or $\dot{\Phi}$) and have vanishing influence on large scales as $k\rightarrow0$ compared to curvature fluctuations.} This makes large-scale observables a powerful probe of the initial conditions that source the large-scale fluctuations in the Universe. Understanding how large-scale fluctuations in the Universe compare to the predictions of $\Lambda$CDM provide insight on the details of the primordial Universe. 

Measuring velocities on large scales is particularly valuable. For example, in many cases the noise associated with the reconstructed velocity fields is constant, making it possible to infer the matter power spectrum with a noise that scales like $k^{2}$ (see e.g.~\cite{Smith:2018bpn}). Since the matter power spectrum can be inferred from galaxies only up to a constant shot noise, this advantage of velocity reconstruction is most important on the largest scales. Of course, inferences on cosmological parameters are still limited by the small number of modes on large scales (cosmic variance). However, one can compare a reconstruction with a galaxy survey to measure bias parameters with arbitrary accuracy; this was proposed in Ref.~\cite{Munchmeyer:2018eey} as a means for detecting primordial non-gaussianity through scale dependent galaxy bias. The moving lens effect provides a measurement of the transverse velocity fields of matter, and has been recently suggested as a tool for cosmological inference in Ref.~\citep{Hotinli:2018yyc}, where the authors introduced a quadratic estimator for the detection of the moving lens effect and reconstruction of transverse velocity fields. An unambiguous detection of the moving lens effect, however, will further benefit from utilising different methods, including using pairwise-velocities~\citep{Yasini:2018rrl}. 

Another method for detecting the moving lens effect is using a matched filter in real space, oriented along the large-scale cosmological bulk velocity. In this work we introduce this method and forecast the detection and reconstruction prospects for the moving lens effect, using a matched filter. This paper is organised as follows: In Section~\ref{sec:moving_lens} we briefly introduce the moving lens effect and the shape of the temperature modulation due to bulk velocities of halos. We calculate the optimal real-space matched filter in~Section~\ref{sec:matchedfilter}. We model the halo and galaxy distribution in Section~\ref{sec:halos_and_so_on}. We discuss the detection prospects for the moving lens effect using these matched filters and halo model in Section~\ref{sec:forecast}. We conclude with discussion in Section~\ref{sec:discussion}. 

\section{The moving lens effect}\label{sec:moving_lens}

Gravitational potentials that evolve in time induce a temperature modulation on the CMB known as the integrated Sachs-Wolfe (ISW) effect which has the form
\be
{ \Theta(\nhat)=-\frac{2}{c^2}\int\frac{\dd \chi}{c}\,\dot{\Phi}(\chi\nhat)}\,,
\ee
where $\Phi(\chi\nhat)$ is the gravitational potential along the line of sight $\nhat$, $\chi$ is the comoving distance, $\Theta=\Delta T(\nhat)/\bar{T}$ {is the fractional CMB temperature fluctuation and we define the integral from the emission of the photon to the observer, unless shown otherwise}. One contribution to the ISW effect in the non-linear regime is the temperature anisotropy due to the peculiar velocity of collapsed structures. This is known as the moving lens effect, and has the form
\be\label{eq:main_1}
\begin{split}
\Theta(\nhat)\!=\!-\frac{2}{c^2}\!\int\!\frac{\dd\chi}{c}\boldsymbol{\nabla}\Phi(\chi\nhat)\cdot\vec{v}_{\rm b,\perp}(\chi\nhat)\,,
\end{split}
\ee
where $\vec{v}_{\rm b,\perp}(\chi\nhat)$ is the peculiar (comoving) transverse bulk velocity. 

\begin{figure}[t!]
    \centering
    \includegraphics[width = 0.6\columnwidth]{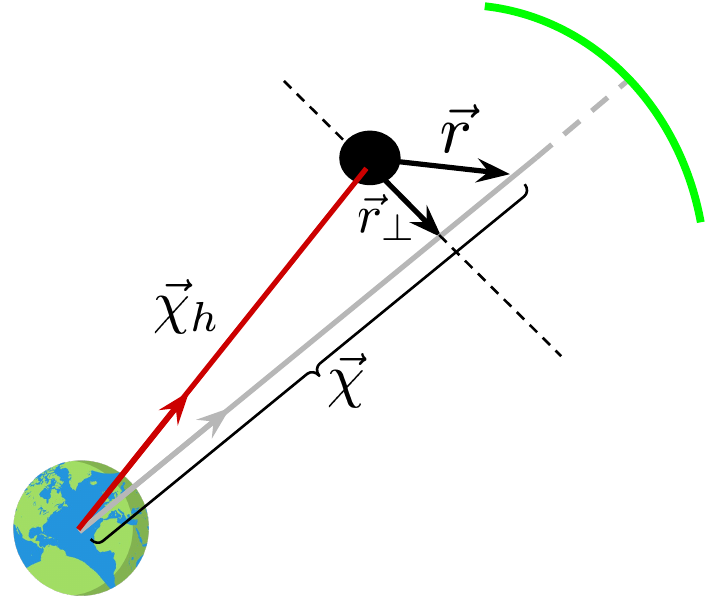}
    \caption{A description of the coordinate system and definitions. We define the comoving distance from the observer on Earth to the DM halo (black circle in the figure) as $\chi_h=|\vec{\chi_h}|$. The comoving distance to the CMB photon is $\chi=|\vec{\chi}|$. Vector $\vec{r}$ connects the halo center to the CMB photon and $r_\perp$ is the transverse distance from the halo center to the trajectory of the observed CMB photon.}
    \label{fig:distances}
\end{figure}

We approximate the {gravitational potential near a halo to be spherically symmetric around the halo center and write}, $\boldsymbol{\nabla}\Phi(r)=\rhat\!~\Phi'(r)$, and $\Phi'(r)=\partial\Phi(r)/\partial r$, {where using Figure~\ref{fig:distances}, we define the unit vector $\hat{\mathbf{r}}=(\vec{\chi}_h-\vec{\chi})/|\vec{\chi}_h-\vec{\chi}|$ and $r=|\vec{\chi}_h-\vec{\chi}|$.} The temperature modulation can then be written as,
\be
\Theta(\nhat)\simeq-\frac{2}{c^2}\int\frac{\dd\chi}{c}\!~\Phi'(r)\!~\left[\rhat\cdot{\vec{v}_{\rm b,\perp}(\chi\nhat)}\right]\,,
\ee
{where the comoving distance $\chi$ depends on $r$ and the distance to the halo.}

We write $\dd\chi=\dd r\,r\,(r^2-r_\perp^2)^{-1/2}$, where $r_\perp=|\vec{r}_\perp|$ and $\vec{r}_\perp$ is the component of $\vec{r}$ orthogonal to the line of sight. The temperature modulation due to moving lens effect takes the form,
\be\label{eq:temp_corr}
\Theta(\nhat)\simeq-\frac{4}{c^3}\left(\vec{v}_{\bb,\perp}\cdot\vec{r}_\perp\right)\int_{r_\perp}^{\infty}\!\!\!\dd r\frac{\Phi'(r)}{\sqrt{r^2-{r}_\perp^2}}\,,
\ee
where $\chi_h\gg r_\perp$ and we approximate the velocity field to be constant within the range of the radial integral, defining a long-wavelength 
(center of mass) bulk-velocity fluctuation as $\vec{v}_{\bb}$ which is the observable we are interested in. There are nevertheless {other non-linear ISW} temperature modulation{s such as the Rees-Sciama effect, for example,} due to the component of the velocity sourced by non-linear growth inside virialized structures (such as clusters), that is uncorrelated with the large-scale bulk flow. While {these} non-linear contribution{s} add to the noise of the velocity measurement on small-scales, we assume {they are} subdominant on large-scales. Note also that the contribution to the moving lens effect from the radial component of the bulk 3-velocity sees $v/c$ relativistic correction when boosted into the CMB frame and is hence sub-dominant, leaving moving lens effect sensitive to the transverse velocities instead.

\begin{figure}[t!]
    \includegraphics[width = \columnwidth]{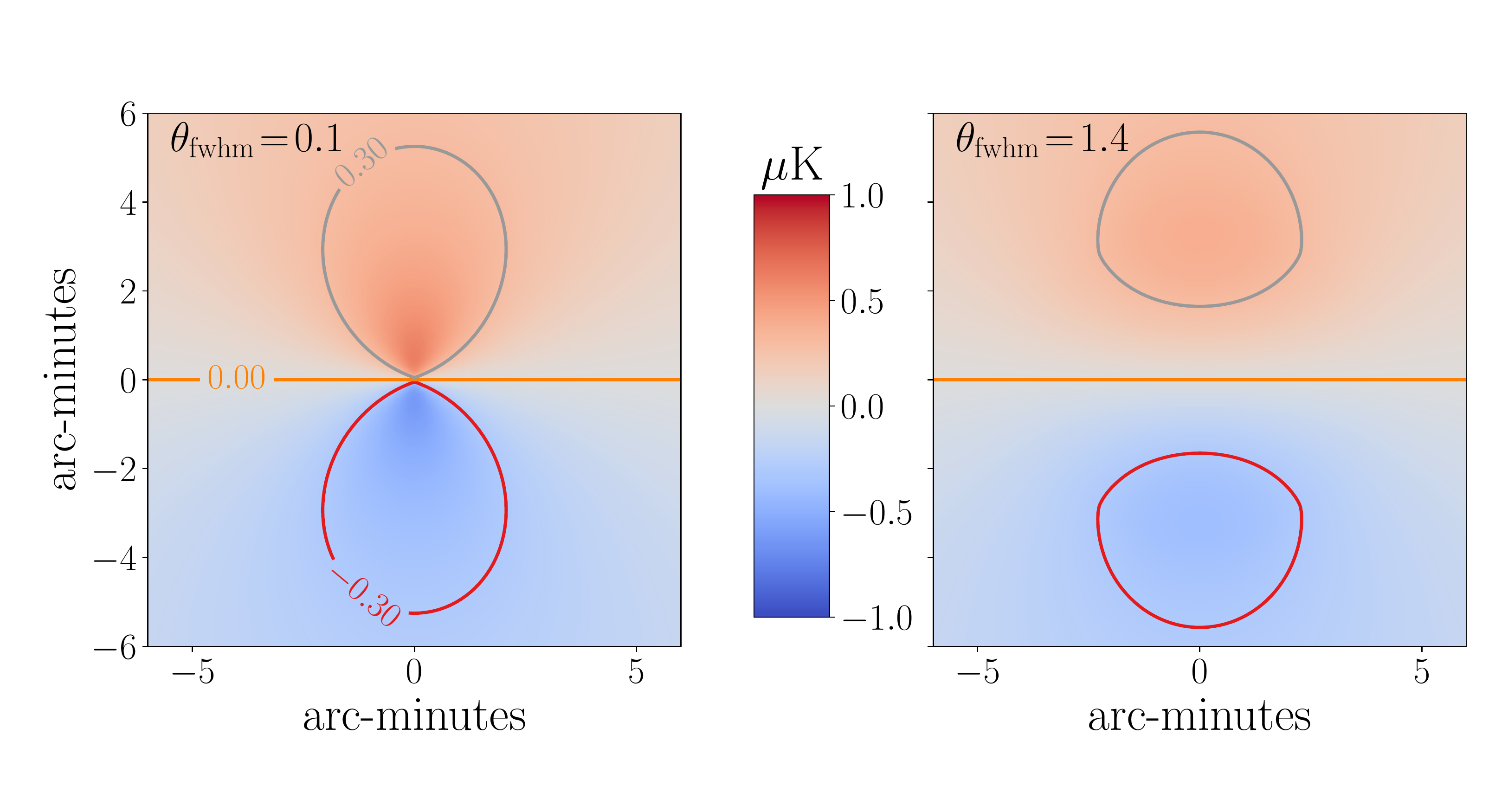}
    \vspace*{-1cm}
    \caption{The CMB temperature modulation due to the moving lens effect shown as a function of comoving radial distance from halo center (in arc-minutes) for an halo of mass $M=10^{14}M_\odot$ {and velocity $v_{b,\perp}=10^{-3}c$,} at redshift $z=1$. The left (right) plots show the templates filtered with a Gaussian beam of radius $0.1$ ($1.4$) arc-minutes.}
    \label{fig:beam_convolve}
\end{figure}

We approximate the functional form of the gravitational potential by using the NFW profile for a spherically symmetric halo with a single parameter, the mass of the halo $M$ in Solar mass units, i.e. $M_\odot\simeq1.989\times 10^{30}~\mathrm{kg}$. We fix the virial radius as 
\be
r_\mathrm{vir}(M,z){\equiv}\left(\frac{G M_\odot M}{100 H^2}\right)^{1/3}\,,
\ee
and assume halos have truncated mass at their virial radius satisfying,
\be\label{eq:mass_cut}
M = \int^{r_{\rm vir}}_0\dd R\,4\pi R^2 \rho(R|M,z)\,,
\ee
where $\rho(R|M,z)$ is the halo profile. The concentration parameter, 
\be
c=A\left(\frac{M}{2\!\times\!10^{12}h^{-1}}\!\right)^\alpha(1+z)^\beta\,,
\ee
relates the scale radius, $r_s(M,z)$, to the virial radius of a halo via $c=r_\mathrm{vir}/r_s$, and we omit showing redshift and mass dependence in what follows. {Note that both scale radius and virial radius are physical distances.} For the model parameters $\{A,\alpha,\beta\}$, we use appropriate values from literature, $\{7.85,-0.081,-0.71\}$. We assume NFW profile for the density of the halo~\citep{Navarro:1995iw},
\be\label{eq:nfw_prof}
\rho(x|M,z)=\frac{\rho_s}{x(1+x)^2}\,,
\ee
and
\be
\Phi(r)=-4\pi G\rho_s r_s^2\frac{\ln(1+x)}{x}\,. 
\ee
{where $x=a r /r_s$ and note that $r$ is the radial comoving distance from the halo center.} We can use the equations above to get 
\be
\rho_s=\frac{M_\odot M}{4\pi r_s^3} \left[-\frac{r_\mathrm{vir}}{r_s+r_\mathrm{vir}}-\ln\left(\frac{r_s+r_\mathrm{vir}}{r_s}\right)\right]\,. 
\ee
The partial derivative of the gravitational potential with respect to $r$ can then be written as 
\be\label{eq:Phi_deriv}
\Phi'(r)=4\pi G\rho_s r_s^2\left[\frac{\ln(1+x)}{x^2}-\frac{1}{x(1+x)}\right]\,. 
\ee
The moving lens signal from a single halo takes the form 
\be
\Theta_{\rm ml}(\vec{x}_\perp)=- a_0 \vec{v}_{\rm b,\perp}\cdot\vec{\mathcal{M}}(\vec{x}_\perp)\, ,
\ee
with 
\be\label{eq:signal_amp}
a_0{\equiv}\frac{16\pi G\rho_sr_s^2}{c^3}\,,
\ee
where $\vec{v}_{\rm b,\perp}$ is  the bulk comoving transverse velocity vector. We calculate the average of the velocity vector components using  \textsc{websky}\footnote{\href{https://mocks.cita.utoronto.ca/data/websky}{Stein~et.al.~\citep{Stein:2020its}} Websky halo catalog covers the full sky up-to redshift $z\sim4.5$, using a mass resolution of  $\sim1.3\times10^{12}M_\odot$.} halo catalog~\citep{Stein:2020its}; as the averaged velocity components of halos inside a volume. The bulk transverse velocity depends on the volume, which we parametrise with the redshift depth and the sky fraction of the patches on the two-sphere. We describe our choices of volume throughout this work. $\vec{x}_\perp=a \vec{r}_\perp /r_s$ and the radial dependence is found by solving Eqn.~\eqref{eq:temp_corr} with Eqn.~\eqref{eq:Phi_deriv} as 
\be\label{eq:mov_lens_temp}
\vec{\mathcal{M}}(\vec{x}_\perp){\equiv}\frac{\vec{x}_\perp}{2x_\perp^2}\!\left[\Big|\frac{2\mathrm{sec}^{-1}(x_\perp)}{\sqrt{x_\perp^2-1}}\Big|+\ln\left(\frac{x_\perp^2}{4}\right)\right]\,,
\ee
where $x_\perp{\equiv}a r_\perp/{r}_s$. The template (shown in Fig.~\ref{fig:beam_convolve}) depends on the mass and redshift of the halo as well as the cosmology through the scale factor.

{\section{The optimal matched filter}\label{sec:matchedfilter}}
We begin by writing the observed real-space intensity map around a dark matter (DM) halo in 2-dimensions as composed of the moving lens signal $\vec{\mathcal{M}}(\vec{r})$ and all other effects 
\be
\Theta^\textrm{obs}({\vec{r}})=-a_0\,\vec{v}_{\rm b,\perp}\cdot\vec{\mathcal{M}}(\vec{r})+\tilde{\Theta}({\vec{r}})\,.
\ee  
We filter our data, $\Theta^\textrm{obs}(\vec{r})$, to get the \textit{unbiased and minimum variance estimate for the components of our bulk transverse velocity signal, projected onto $\{\hat{x},\hat{y}\}$ Cartesian orthogonal directions, $\hat{v}_{b,\perp,x}$ and $\hat{v}_{b,\perp,y}$,} which we define as 
\be\label{eq:signal_estimate}
\hat{v}_{\bb,\perp,i}{\equiv} a_0^{-1}\!\!\int \dd^2{\vec{r}}\,\Psi_i({\vec{r}})\Theta^{\rm{obs}}({\vec{r}})\,.
\ee
where $i\in\{x,y\}$. (Please see an alternative method for estimating the norm and the angle of the velocity vector and the velocity in Appendix~\ref{sec:polar_estimator}.)
The transverse velocity amplitude is degenerate with the density and the scale radius of the halo, which are determined by halo mass and redshift. We comment on these degeneracies in the following sections. 

The observed fractional intensity maps satisfy 
\be
\langle\tilde{\Theta}(\vec{\ell})\rangle\!=\!0\ \ {\rm{and}}\ \ \langle\tilde{\Theta}(\vec{\ell})\tilde{\Theta}(\vec{\ell}\,')\rangle\!=\!(2\pi)^2\delta(\vec{\ell}\!+\!\vec{\ell}\,')\tilde{C}_\ell^{TT} \, , \non
\ee
where $\tilde{C}_\ell^{TT}$ is the lensed CMB temperature power spectrum including idealized experimental noise and the kSZ contribution, which we approximate as a constant $3~\mu K^2$ in $\ell(\ell+1)/(2\pi)^2C_\ell^{\Theta\Theta}$. We assume perfect removal of foregrounds such as the cosmic infrared background and tSZ from the CMB.
Note also that we assume noise and foregrounds can be approximated by Gaussian random fields. This neglects statistical anisotropies e.g. due to weak lensing of the CMB, which induces a similar dipole pattern around halos aligned with the large-scale CMB gradient, biasing the moving lens measurement. As shown in Ref.~\cite{Hotinli:2018yyc}, however, the ordinary lensing bias remains negligible on scales $\ell\lesssim50$, or for patches larger than $\sim2$\,degrees in radius. CMB lensing also boosts the variance of the CMB sampled near galaxies, which we discuss in Section~\ref{sec:discussion}. 

We define the matched filter such that the estimator recovers the true velocity, and define parameters $b_i{\equiv}\langle\hat{v}_{\bb,\perp,i}-{v}_{\bb,\perp,i}\rangle$ and ${N}^{\rm rec}_i{\equiv}\langle(\hat{v}_{\bb,\perp,i}-{v}_{\bb,\perp,i})^2\rangle\,,$ where
\be
{b_i{\equiv}\int\!\dd^2\vec{r}\,\Psi_i(\vec{r}){\mathcal{M}_i}(\vec{r})-1\,,}
\ee
and
\be
\begin{split}
N_i^{\rm rec}&=a_0^{-2}\!\!\int \frac{\dd^2\vec{\ell}}{(2\pi)^2} {|\tilde{\Psi}_i(\vec{\ell})|^2}{\tilde{C}_\ell^{TT}}\,. 
\end{split}
\ee
We defined the Cartesian projections of our template as $\mathcal{M}_x(\vec{x}_\perp) = \vec{\mathcal{M}}(\vec{x}_\perp)\cdot\hat{x}$ and $\mathcal{M}_x(\vec{y}_\perp) = \vec{\mathcal{M}}(\vec{x}_\perp)\cdot\hat{y}$. 

We now wish to minimize the variance of our filter under the condition that the bias vanishes. We do this by defining $\mathcal{L}{\equiv}N^{\rm rec}_i+\lambda b$ where $\lambda$ is now a Lagrange multiplier and 
\be
\mathcal{L}\!=\!\int\frac{\dd^2\vec{\ell}}{(2\pi)^2} \tilde{\Psi}_i^\star(\vec{\ell})\left[a_0^{-2}\tilde{\Psi}_i(\vec{\ell})\tilde{C}_\ell^{TT}+\lambda\tilde{\mathcal{M}}_i(\vec{\ell})\right]-\lambda\,.\ \ \ \
\ee
The optimal filters that minimize $\mathcal{L}$ can be written as  
\be
\tilde{\Psi}_i(\vec{\ell})=\left[\int\frac{\dd^2\vec{\ell}'}{(2\pi)^2}\frac{|\tilde{\mathcal{M}}_i(\vec{\ell}')|^2}{\tilde{C}_{\ell'}^{TT}}\right]^{-1}\frac{\tilde{\mathcal{M}}_i(\vec{\ell})}{\tilde{C}_{\ell}^{TT}}\,,
\ee
or equivalently, 
\be
\tilde{\Psi}_i(\vec{\ell})={N}^{\rm rec}\frac{\tilde{\mathcal{M}}_i(\vec{\ell})}{\tilde{C}^{TT}_\ell}\,, 
\ee
where we defined $N^{\rm rec}\equiv N^{\rm rec}_i$ and used $N^{\rm rec}_x=N^{\rm rec}_y$.

Note that the optimal estimators are most sensitive to the signal on small scales, where the inverse of the estimator variances are large and the primary CMB signal (which is much larger and acts as a confusion) is small. Lastly, we convolve the moving lens signal with a beam that matches the experimental specifications described below. When applying the matched filter we assume a Gaussian beam satisfying  $B(\vec{\ell})=\exp[-({\rm \theta_{\rm fwhm}}/2\sqrt{\ln2})^2\ell(\ell+1)]\,,
$  where $\theta_{\rm fwhm}$ is the full beam-width at half-maximum. In what follows we discuss results with this beam applied to the moving lens templates, i.e. $\tilde{\mathcal{M}}_i(\vec{\ell})\rightarrow B(\vec{\ell})\tilde{\mathcal{M}}_i(\vec{\ell})$. \\ 

\section{Halos, galaxies and the $\hat{v}_{\bb,\perp,i}$ SNR}\label{sec:halos_and_so_on}

The estimated signal-to-noise ratio (SNR) for the velocity amplitude per object with mass $M$ at redshift $z$ is $[\sum_i{v}^2_{\bb,\perp,i}/N^{\rm rec}_i(M,z)]^{1/2}$; the number of such objects needed for total SNR to equal to 1 is $[\sum_i{v}^2_{\bb,\perp,i}/N^{\rm rec}_i(M,z)]^{-1}$. While upcoming  surveys will not be able to reconstruct the transverse velocity for each halo, the average transverse velocity can be measured over a sufficiently large patch of the sky.

Surveys of large-scale structure observe galaxies that occupy DM halos. The relation between galaxies and the host DM halos depend on a multitude of effects and mechanisms, including rates of star formation and galaxy mergers, and needs to be modelled and tested against data. The number and spatial distribution of the DM halos can be described by the halo model (see for review e.g.~\citep{Cooray:2002dia}). The distribution of galaxies inside DM halos can be described with a halo occupation distribution (HOD) model (see e.g.~\citep{2011ApJ...738...45L}) where every DM halo is assumed to have at most 1 central galaxy, as well as additional satellite galaxies whose number can be large for massive halos. The average observable central (satellite) galaxy number count of a DM halo with mass $M$ and at redshift $z$ is parametrised with $\bar{N}_c(m_*,z)$ [with $\bar{N}_s(m_*,z)$] where $m_*$ is the threshold stellar mass determined by the galaxy survey and details of the model can be found in e.g.~\cite{2011ApJ...738...45L,2012ApJ...744..159L,Berlind:2001xk}. For  calculating the mass and redshift dependence of halo density we assume a Sheth-Tormen collapse fraction~\citep{Sheth:1999su}. 

We use the matched filter introduced above and the halo mass function with a normalisation appropriate for a given LSS survey and approximate the expected total ${\rm SNR}^2=\sum_i {\rm SNR}_i^2$ for the velocity magnitude from inside a redshift bin and a given patch of size $f_{\rm sky}^{\rm patch}$ on the sky with
\be
\begin{split}
{\rm S}&{\rm NR}^2_i =4\pi f_{\rm sky}^{\rm patch} \\ & \times \!\!\!\!\int\limits_{z-\rm bin}\!\!\!\!\!\dd z\!\!\!\!\!\!\!\!\int\limits_{\ \ \ \ \rm catalog}\!\!\!\!\!\!\!\!\dd M\bar{N}_c(m_*,z)\frac{v_{\bb,\perp,i}(z)^2}{{N_i^{\rm rec}(M,z)}}\chi^2\frac{\dd \chi}{\dd z}n(M,z)\!\,.
\end{split}\non
\ee
Note that we use only the count of central galaxies, since the bulk transverse velocity is sourced by the center of mass of the halo.
{We find ${\rm{SNR}}^2\!\simeq\!10^{3}f_{\rm sky}$ for a redshift bin centered at ${z}=1$ and of size $\Delta z=1$}, with perfect knowledge of halo mass, location and redshift as well as the transverse velocity direction, using Vera.~C.~Rubin Observatory and CMB-S4 experimental specifications using the analytic approximation for the galaxy number density satisfying $\dd n/\dd z\propto(z/z_0)^\alpha\exp[(-z/z_0)^\beta]{\rm arcmin}^{-2}$ with $\{z_0,\alpha,\beta,n_{\rm tot}[{\rm arcmin}^{-2}]\}$ set equal to $\{0.3,2,1,40\}$, and the CMB temperature noise $N_\ell^{TT}=(\Delta_T/T)^2\exp{[\ell(\ell+1)\theta_{\rm fwhm}^2/(8\log(2))]}$ where we set $\{\Delta_T,\theta_{\rm fwhm}\}$ to $\{1.0,1.4\}$. Note that individual halo masses are expected to be measured imperfectly, with around $40$ percent error in $\ln M$, from combinations of lensing and SZ measurements and redshift measurements are subject to photo-$z$ errors~\citep{Palmese:2019lkh,Murata:2017zdo,Ballardini:2019wxj}. We discuss these in Section~\ref{sec:forecast}, before forecasting on the transverse velocity amplitude  reconstruction fidelity of the upcoming surveys in cross correlations of CMB and galaxy measurements.  

\section{Reconstruction, Cross-correlations and Forecasts}\label{sec:forecast}

We evaluate the detection SNR on the moving lens effect for a given patch and a redshift range as the sum of the SNR on the two transverse velocity components we reconstruct, ${v}_{\bb,\perp,x}$ and ${v}_{\bb,\perp,y}$; and the total SNR \textit{per patch} as sum SNR of the components, ${\rm SNR}^2\equiv{\rm SNR}^2_x+{\rm SNR}^2_y$.
When estimating the total detection SNR from the full sky we assume no correlation between patches and set ${\rm SNR}_{\rm total}^2\simeq f_{\rm sky}/f_{\rm sky}^{\rm patch} {\rm SNR}^2$ where $f_{\rm sky}$ is the full sky coverage of the cosmological survey. We calculate the SNR in volumes of redshift depth $\Delta z=0.5$ in the range $z\in[0.1,3]$ and surface area corresponding to the patch size. 

\begin{figure}[t]
    \centering
    \includegraphics[width = \columnwidth]{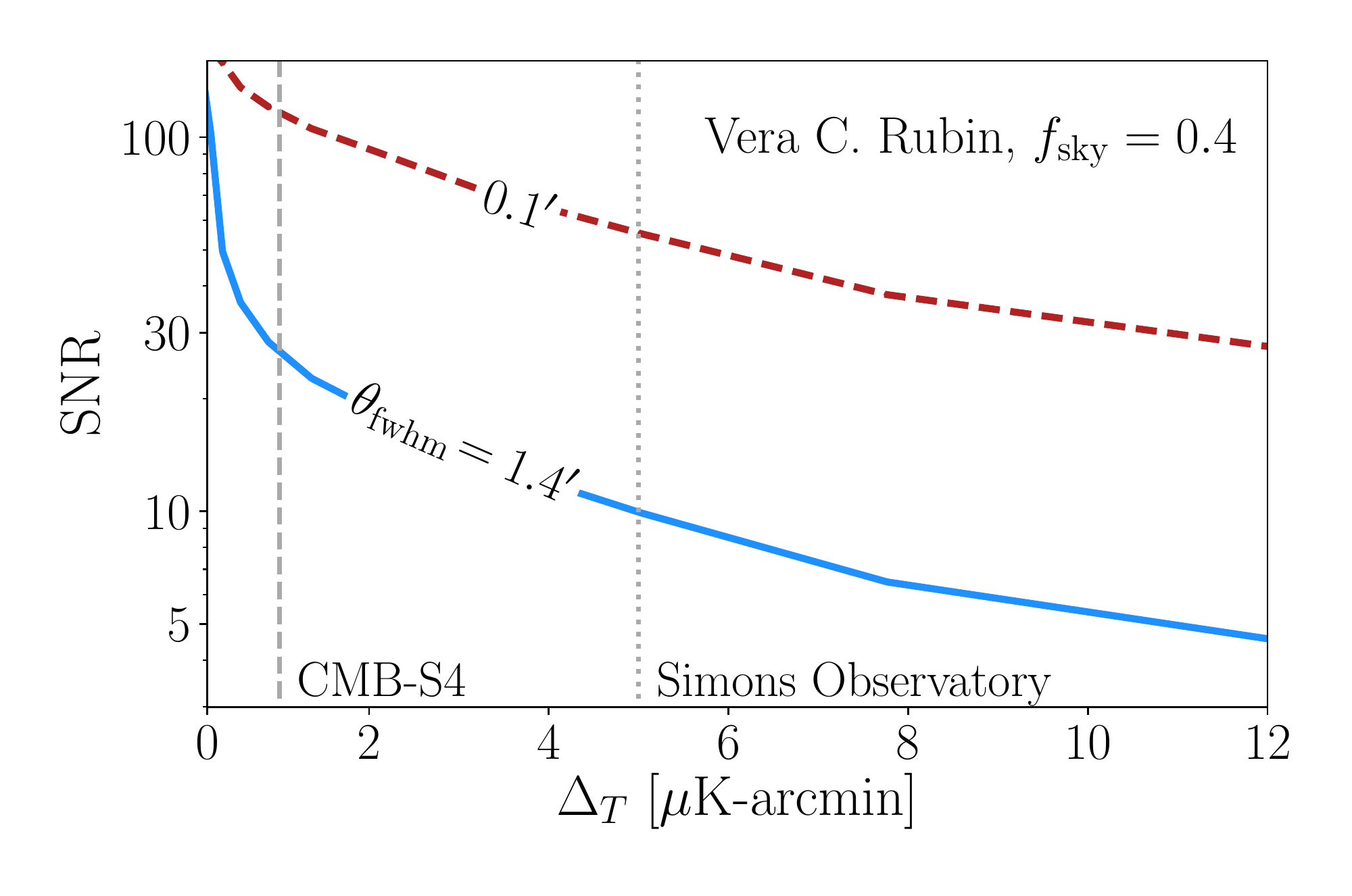}
    \vspace*{-0.75cm}
    \caption{The total transverse-velocity detection SNR from the measurement of the moving lens effect, for 1.4- and 0.1-arcminute beams for various CMB rms noise levels. Plotted curves show SNR values for halo counts matching with the expected central galaxies from LSST survey (a.k.a~Vera~C.~Rubin Observatory) and a sky fraction of $f_{\rm sky}=0.4$.     \label{fig:snr_all}}
    \vspace*{-0.5cm}
\end{figure}

We display the forecasts for total SNR for moving lens effect detection in Figure~\ref{fig:snr_all} using patches of area 16 square degrees. Our calculation suggest the upcoming surveys of LSS and measurements of CMB may detect the moving lens signal to high significance, where combinations of CMB-S4 and Vera~C.~Rubin Observatory will achieve SNR of around 20 and combinations of SO and Vera~C.~Rubin Observatory will achieve SNR of around 10. These results are consistent with the results obtained using the quadratic estimator of Ref.~\citep{Hotinli:2018yyc}.

In reality, imperfect modelling of the halo profile or the background cosmology, along with potential systematic errors in the halo mass and the halo redshift measurements, may bias the estimator of the true velocity field from the moving lens effect, yielding $\langle\hat{v}_{\bb,\perp,i}^{\rm ml}\rangle=b_{\rm ml}\,{v}_{\bb,\perp,i}$. This is analogous to the optical depth degeneracy encountered when attempting to reconstruct the radial velocity field using kinetic Sunyaev Zel'dovich (kSZ) tomography (see Ref.~\cite{Smith:2018bpn} for a discussion). Furthermore, -- even in the absence of potential biases -- the imperfect knowledge of these parameters leads to additional uncertainly and hence boosts the variance of the velocity estimator.

Imperfect knowledge of the halo mass effects the fidelity of the velocity measurement due both to a reduced filter response and the intrinsic degeneracy between the halo mass and true velocity amplitude. The velocity estimator is proportional to $(M/M_\odot)^{-0.6}$, as defined in Eq.~\eqref{eq:signal_estimate}. 
In order to evaluate the unambiguous detection and reconstruction significance of the upcoming experiments on the transverse velocity amplitude, we have to incorporate the error in halo masses in the velocity SNR calculation. The error on halo mass is expected to satisfy (per halo) $\sigma_{M}/M\simeq0.4$, using mass-richness measurements from weak lensing and SZ surveys~\citep{Murata:2017zdo,Palmese:2019lkh}. Note that this mass error is already significantly more optimistic than the moving lens SNR per halo; which satisfy (per halo) $\sigma_{v_{\bb,\perp,i}}/|{v_{\bb,\perp,i}}|>1$ for all of the observable redshift and halo mass ranges. Hence we find that the error induced on the velocity SNR due to halo mass degeneracy to be over $\mathcal{O}(10)$ smaller than the error on the moving lens amplitude $A$. In order to evaluate the reduction in SNR due to mass errors on the template, we calculated moving-lens temperature maps $\Theta_{\rm ml}(\nhat)$, as defined above, using \textsc{websky} halo catalog~\citep{Stein:2020its} with the true and erroneous halo masses, with the latter having random halo masses sampled from a Gaussian distribution with $\sigma_M/M\simeq0.4$. We find the cross-correlation coefficient of the moving-lens temperature maps remains near $1$ for the multipole ranges relevant to this study ($\ell\lesssim100)$ -- suggesting the errors induced by the inaccurate template are small on large scales, and furthermore, that the cross-correlations with an external tracer of the density field may be used to potentially boost the SNR. 

{Another important parameter that can penalize the SNR is the halo redshift, since the velocity reconstruction from galaxy-surveys suffer from known effects of redshift space distortions (RSDs) as well as photometric redshift (photo-$z$) errors for photometric surveys such as the Vera~C.~Rubin Observatory, the latter satisfying $\sigma_z=0.03(1+z)$. Similar to the uncertainty on halo masses, we find the contribution to the error on the velocity measurement due to the redshift degeneracy (induced by dependence of the signal on the scale-radius) to be small, especially since we use large ($\Delta z\sim0.5$) redshift bins. In order to evaluate the significance of photo-$z$ errors and RSDs on matched filter, we sample from the same halo catalog, a smaller set of halos with halo number count matching the expected central galaxies from the Vera~C.~Rubin Observatory. We compare the true velocity fields with and without taking into account the photo-$z$ errors and RSDs. We capture the effect of photo-$z$'s by redistributing the halo positions in redshift space randomly with the photo-$z$ error. For RSDs, we add the velocity dependent RSD correction in redshift space, as discussed above. We parametrise the combined effect of RSDs and photo-$z$ with the cross-correlation coefficient $\varrho=C^{xy}_\ell/\sqrt{C_\ell^{xx}C_\ell^{yy}}$ as a function or redshift, where $\{x,y\}\equiv\{{\rm true},{\rm obs}\}$, which we find remains larger than 90 percent for $\ell\lesssim100$ -- showing (similarly to the mass errors) that the redshift errors do note degrade the SNR from the moving lens measurement and that the velocity field measured from the templates is well correlated with the underlying velocity field.}

We leave a more detailed analysis of possible contribution to noise and biases from other effects including other CMB secondaries to an upcoming work. 

Note that the transverse velocity fields vary more rapidly along their projected direction and a similar phenomena is also true for the radial velocities, i.e. they vary more rapidly on the radial direction. This suggests that the typical transverse velocity modes vary slower in the radial direction, implying the relative SNR penalty from larger redshift bins (necessitated by the large photo-$z$ errors from photometric surveys) is lower than compared to radial velocity reconstruction for the kSZ effect, for example, motivating the use of photometric surveys for the purpose of moving lens effect detection and velocity reconstruction. Note however that the cross-correlation coefficient suffers due to low number of galaxies in the near Universe ($z<0.3$) suggesting potential benefits of using different types of observations (such as spectroscopic surveys and other tracers) for the purpose of moving-lens effect detection. We leave a more involved analysis on these lines to future work.

\section{Discussion}\label{sec:discussion}

In this paper, we have shown that the dipolar pattern in the CMB temperature fluctuations around moving DM halos due to the moving lens effect can potentially be detected in the near future using a matched filter in real space. These real space filtering techniques can potentially be used to reconstruct the bulk velocity fields in the Universe.  We calculated the form of the optimal matched filter, which is imagined to be centered on DM halos inferred from galaxy surveys, and aligned with the cosmological bulk transverse velocities. We discuss the distribution and the detection prospects of halos from galaxy surveys, as well as the effect of photo-$z$ errors and redshift-space distortions on the bulk velocities inferred from a halo catalog. We calculate estimates for the SNR with the upcoming experiments using analytic expressions we derive. We find that a statistically significant detection will be possible with the Simons Observatory, upon cross-correlation with Rubin galaxy survey, for example.    

The maximum residual signal resulting from stacking a large number of halos inside volumes of size around the correlation length of the cosmological velocity field can potentially be used to estimate the direction and amplitude of the bulk velocity at a given region. Using the known functional form of the moving lens effect could potentially increase the accuracy of reconstruction by fitting the template calculated Eqn.~\eqref{eq:mov_lens_temp}. In this study, we introduce a velocity reconstruction technique from applying a real-space matched filter and oriented CMB stacking. Measurements of large-scale velocity fields provide an effective probe of the early Universe signatures such as non-Gaussianity (e.g.~\citep{Munchmeyer:2018eey}) as well as the absolute growth rate, which can be constraining for studies of dark energy models, modified gravity and effects of neutrino mass. Once detected, transverse velocity fields reconstructed from measurement of the moving lens effect can potentially be useful for studying variety of interesting signatures and models.

Transverse velocity modes also provide a rare window into measuring the profile of DM halos and can afford constraining power on various halo model parameters upon cross-correlating with other tracers of large-scale structure such as weak gravitational lensing. We leave a more detailed study on the prospects of testing halo models with the moving lens effect to a future work. 

Note that another velocity-dependent effect on the CMB is the so-called `rotational' kSZ effect due to the rotational motion of the large galaxy clusters~\citep{Cooray:2001vy,Chluba:2002es}, sourced by the angular momentum of halos.
Various recent studies~\citep[e.g.][]{Matilla:2019yhu, Baxter:2019tze} show that ongoing experiments will have the statistical power to detect this dipolar kSZ signature centered around galaxy clusters. On small scales, the dominant contribution to the rotational kSZ is the component of the angular momentum field that is sourced by the non-linear growth and dynamics of the virialized environment, which is not correlated with bulk transverse velocity. This component acts as noise on the moving lens measurement. Nevertheless, the angular momentum field is not expected to be entirely uncorrelated with the long wavelength potential: correlations are induced due to deviations of the proto-halos from perfect spherical symmetry and their alignments, for example. The correlated rotational kSZ effect can bias the moving lens measurement as well as potentially providing information regarding the growth of structure and the initial conditions. We leave a detailed analysis on the implications of large-scale angular momentum correlations on the moving lens and kSZ effect measurements to future work.

Finally, since photons traversing near galaxies trace regions of the Universe with larger density fluctuations, patches we use in our real-space analysis are more noisy in average than a random location on the sky due to ordinary lensing. The effect of ordinary lensing on the CMB is not random, however, and the induced dipolar pattern is correlated with the large-scale CMB gradient. Since the CMB gradient is very well measured, this correlated boost in variance (which can be understood as a bias induced by ordinary lensing) can potentially be removed. Since the temperature gradient is not correlated with the bulk transverse velocity, we anticipate this procedure will not degrade the detection prospects of the moving lens effect significantly. Note that we also checked for the additive noise effect from (nonphysical) randomly distributed ordinary lensing contribution to the CMB spectra (which is a shot-noise term, $1/N_{\rm source}$, on the templates where $N_{\rm source}$ is the number of sources available from the galaxy catalogue). For the upcoming LSS surveys, together with Simons Observatory and CMB-S4, for example, we find this shot noise remains smaller compared to the noise of the CMB experiment.

In the next years, measurements of secondaries will become observationally significant for the first time as CMB and LSS surveys will achieve the necessary precision. Precision measurement of small scale CMB fluctuations will open new windows of opportunities for cosmological inference. In this work we discussed the prospects of moving lens effect detection from upcoming surveys with a real-space matched filter. Complementary to quadratic estimator technique introduced in~\citep{Hotinli:2018yyc}, the real-space analysis provides a useful alternative which will allow high SNR detection of the moving lens effect in the near future.  

\section{Acknowledgements}

We thank Jim Mertens for collaborations in the early stages of this work. We thank Simone Ferraro, Colin Hill, Andrew Jaffe, Yinzhe Ma, Moritz Munchmeyer, Emmanuel Schaan and David Spergel for helpful discussions. We thank an anonymous referee for comments that improved the derivation of the optimal filter. SCH is supported by the Perimeter Institute Graduate Fellowship and the Imperial College President's Scholarship. MCJ is supported by the National Science and Engineering Research Council through a Discovery grant. JM is supported by the US Department of Energy under grant no.~DE-SC0010129. This research was supported in part
by Perimeter Institute for Theoretical Physics. Research at Perimeter Institute is supported by the Government of
Canada through the Department of Innovation, Science and Economic Development Canada and by the Province
of Ontario through the Ministry of Research, Innovation and Science.

\bibliography{main}

\appendix

\section{Polar estimator}\label{sec:polar_estimator}

In this section we introduce an alternative reconstruction to the Cartesian velocity-component estimator we defined above. While the estimator introduced here is (marginally) less optimal, it can be used for validation purposes. We define the filter response introduced in Eq.~\eqref{eq:signal_estimate} for a halo `$i$' of mass $M_i$ and redshift $z_i$, as 
\be
\hat{A}(M_i,z_i){\equiv}\int\dd^2\vec{r}\,\Psi_i (\vec{r})\Theta^{\rm obs} (\vec{R}_i+\vec{r})\,,
\ee 
where $\Theta^{\rm obs}(\vec{R}_i\!+\!\vec{r})$ is the observed CMB around the halo at $\vec{R}_i$ (in polar coordinates) from the patch center $\vec{R}_0=(0,0)$. The matched filter centered on the halo, $\Psi_i(\vec{r})$, depends on the halo mass and redshift as well as the orientation of the transverse velocity field, which we assumed known in the previous section. In this section we evaluate the prospects for reconstructing the components of the transverse velocity vector from the CMB and halo locations from a galaxy survey.

We are interested in finding the the angle $\hat{\varphi}_0$ that best approximates the true average angle of the transverse velocity vector field with respect to a reference vector in a patch of size $4\pi f_{\rm sky}^{\rm patch}$, where we set $\hat{\varphi}_{0,i}=\hat{\varphi}_{0,j}=\hat{\varphi}_0$ equal for all filters $\{i,j\}$ inside the patch. This is the $\hat{\varphi}_0$ that satisfies\footnote{We assume the maxima can be distinguished from the minima from the filter response and accounted for with a sign change, with no additional error to the estimator.},
\be\label{eq:derivative}
\int\dd^2\vec{r}\frac{\partial}{\partial\varphi_0}\sum_{i,\rm halos} \Psi_i(\vec{r})\Theta^{\rm obs}(\vec{R}_i+\vec{r})=0\,,
\ee
where for each filter in the sum, the coordinates are chosen so that the halo is at the center of the template.The CMB acts as noise on the stacked patch, and that the $\varphi_0$ which maximizes the residual response approximates the true direction of the transverse velocity vector direction on 2-sphere, given sufficient SNR. 

Due to the simple angular dependence of the signal profile, we find the equality in Eqn.~\eqref{eq:derivative} satisfies, \be
\tan(\hat{\varphi}_0,z)\!=\!\frac{\int\dd^2\vec{r}\sin\varphi\sum_i{\Psi_{i,u}}(r)\Theta^{\rm obs}(\vec{R}_i+\vec{r})}{\int\dd^2\vec{r}\cos\varphi\sum_i{\Psi_{i,u}}(r)\Theta^{\rm obs}(\vec{R}_i+\vec{r})},\ \ 
\ee where we define $\Psi_i(r)\!=\!\cos(\varphi\!-\!\hat{\varphi}_0)\Psi_{i,u}(r)$, $\hat{\mu}{\equiv}\hat{\alpha}/\hat{\beta}\!=\!\tan\hat{\varphi}_0$ (omitting showing the redshift dependence for now); where $\alpha$ and $\beta$ are the numerator and denominator on the right-hand-side of Eq.~\eqref{eq:derivative}, respectively. The error on the measurement can be written in the form 
\be\sigma_\mu/|\mu|= \sqrt{\sigma^2_\alpha/\alpha^2+\sigma^2_\beta/\beta^2}\,.
\ee

We assume the contribution to the \textit{signal} from all else except moving lens effect vanish for a large enough patch with sufficiently many halos. With this assumption, we write\footnote{Similarly equality holds for $\beta$, with azimuthal angular integral over $\cos\varphi$ in-place of $\sin\varphi$.}
\be\label{eq:signal__}
\begin{split}
\alpha&=\!\!\!\!\!\!\int\limits_{\rm template}\!\!\!\!\!\dd^2\vec{r}\sin\varphi\!\!\sum\limits_{i,\rm halos}\!\!{\Psi_{u,i}}(r)\Theta^{\rm ml}(\vec{r})\,,
\end{split}
\ee
where we defined $\Theta^{\rm ml}(\vec{r}){\equiv}A(M,z) \!~\mathcal{M}_u(r)\cos(\varphi\!-\!\varphi_0)$, $\mathcal{M}_u(r)$ is the radial shape of the moving lens effect on the CMB around a halo where $r$ is the radial distance to the halo center, {and we defined the (polar) integral over the patch as equal to the surface area of the patch on 2-sphere, as $\int_{\rm patch}\dd^2\vec{\Omega}\simeq\int_{\rm patch}\dd^2\vec{R}{\equiv}4\pi f_{\rm sky}^{\rm patch}$} for small patches. A more detailed derivation can be found in Appendix~\ref{sec:appendix_1}. Errors are  calculated using a relation similar to Eq.~\eqref{eq:signal__}, with the CMB component without the moving lens effect instead, and performing the average over the realisations of the CMB as e.g.$^{3}$,
\be\label{eq:noise}
\begin{split}
\sigma_\alpha^2=\!\!\!\!\!&\!\!\!\int\limits_{\rm template}\!\!\!\!\!\!\!\dd^2\vec{r} \sin\varphi\,\dd^2\vec{r}\,'\sin\varphi'\,\\
&\times\!\!\!\!\sum_{i,j,\rm halos}\!\!\!\!\Psi_{i,u}(r)\Psi_{j,u}^*(r')\langle\tilde{\Theta}(\vec{R}_{i}\!+\!\vec{r})\tilde{\Theta}^*(\vec{R}_{j}\!+\!\vec{r}\,')\rangle\,.
\end{split}\non
\ee

For compactness of our expressions we define a signal parameter $I_{\rm patch}{\equiv}\alpha/(\pi\sin\varphi_0)\!=\!\beta/(\pi\cos\varphi_0)$ which satisfy, \be\label{eq:azimuth_est}
\begin{split}
I_{\rm patch}&
=4\pi f_{\rm sky}^{\rm patch} \\ &\times \!\!\!\int\limits_{{z}-\rm bin}\!\!\!\!\dd z\,\!\!\!\!\!\!\!\!\!\int\limits_{\rm \ \ \ catalog}\!\!\!\!\!\!\!\!\dd M \bar{N}_c(m_*,z)\, \chi^2\frac{\dd \chi}{\dd z}n(M,z)A(M,z)\,, \end{split}\non
\ee where we set the integral over the patch in Eq.~\eqref{eq:signal__} as $\int_{\rm patch}\!\dd^2\vec{R}{\equiv}4\pi f_{\rm sky}^{\rm patch}$ and like before, we promote the sum over halos to an integral over halo masses and the halo locations over the patch. The error on $I_{\rm patch}$ takes the form,
\be
\begin{split}
\sigma_{I_{\rm patch},\bar{z}}^2&=\!\!\!\iint\limits_{\ \ \ r_{\rm min}}^{\ \ \ r_{\rm max}} \!\! r\dd r \, r'\dd r'\Lambda(r,r') \mathcal{F}\!_{\bar{z}}(r)\mathcal{F}_{\bar{z}}^*(r')\,,
\end{split}
\ee
where we defined 
\be
\mathcal{F}\!_{\bar{z}}(r){\equiv}\!\!\!\int\limits_{{z}-\rm bin}\!\!\!\!\dd z\,\!\!\!\!\!\!\!\!\int\limits_{\rm \ \ \ catalog}\!\!\!\!\! \!\!\! \dd M \bar{N}_c(m_*,z)\, \chi^2(z)\frac{\dd \chi}{\dd z}n(M,z)\Psi_{u}(r)\,,\non
\ee
and
\be
\begin{split}
\Lambda (r,r'){\equiv}&16\pi^4\!\!\int \frac{L^{-1}\dd L}{(2\pi)^2}\,
C_L^{\tilde{\Theta}\tilde{\Theta}}J_1(L r)J_1(L r')\\
&\,\,\times \left[R_{\rm max}J_1( R_{\rm max})\!-\!R_{\rm min}J_1( R_{\rm min})\right]^2\!, 
\end{split}\non
\ee
where $r_{\rm max}$ satisfies the inequality ${r_{\rm max}\!\ll\!R_{\rm max}}$. We set $\{r_{\rm max},r_{\rm min},R_{\rm min}\}$ equal to $\{5\,{\rm arcmin},1.4\,{\rm arcmin}, r_{\rm max}\}$ and find $\Lambda(r,'r')\simeq\mathcal{A}rr'$, where  $\mathcal{A} \!\simeq\!2.3\!\times\!10^{-11}$ for $R_{\rm max}\!=\!2\!\times\!10^{-2}$radians and $\mathcal{A}$ depends on $R_{\rm max}$ non-trivially due to the scale dependence of the CMB. This term can be understood as the r.m.s.~contribution of the CMB on the noise estimate for a given patch, and is independent from the CMB noise to a good approximation for current and upcoming CMB experiments with the sufficiently large $r_{\rm min}$ choice we make above. Generally the integral limits $\{r_{\rm min},r_{\rm max}\}$ can be chosen as halo mass dependent to maximize the SNR. Using these relations we get 
\be
\sigma_{I_{\rm patch}}^2 \simeq \mathcal{A} \,\Big|\!\int\!\frac{ \dd \ell}{2\pi}\!\!\!\!\!\!\int\limits_{\rm catalog}\!\!\!\!\!\dd M\zeta_\ell(M,z) \tilde{\mathcal{F}}_{(\bar{z})}'(\ell)\Big|^2\,,
\ee
where 
\be 
\zeta_\ell(M,z){\equiv}\int\limits_{r_{\rm min}}^{r_{\rm max}}r^2 \dd r \exp(-i\ell r)\,,
\ee 
and $\tilde{\mathcal{F}}'(\ell){\equiv}\dd \tilde{\mathcal{F}}(\ell) / \dd M$ where $\tilde{\mathcal{F}}(\ell)$ is the (1D) Fourier transform of $\mathcal{F}(r)$. Note it is straight-forward to show from equations above that $I_{\rm patch}$ satisfies the equality, 
\be
\sigma_{\mu}/|\mu|=\sqrt{2}\sigma_{I_{\rm patch}}/|I_{\rm patch}|\,,
\ee 
in the perfect knowledge of the moving-lens amplitude $A$.\footnote{Note that the error on the amplitude $A$ can be added to give \be\label{eq:footnote} \sigma_\mu^2/\mu^2=2(\sigma_{I_{\rm patch}}^2/I_{\rm patch}^2+\sigma_A^2/A^2)\,.\ee } 

We evaluate the detection significance of the direction component $\mu$ (ignoring the uncertainty on the amplitude) using our parameter choices, for a Rubin-like halo catalog and a CMB-S4-like survey, and 6 (uncorrelated) boxes equally spaced in redshift in the range $z\in[0.1,3]$ with same surface area on the sky. We find ${\rm SNR}\gtrsim1$ for a patch with surface area of $\lesssim10$ square degrees. The detection SNR on the moving lens effect for a given patch and a redshift range can be evaluated as the sum of the SNR on the two transverse velocity components we reconstruct from the velocity amplitude and the \textit{angle} as 
\be
\vec{v}_{\bb,\perp}{\equiv}\{v_1,v_2\}=\{v_{\bb,\perp}\cos{\varphi_0},v_{\bb,\perp}\sin{\varphi_0}\}
\ee 
and the total SNR \textit{per patch} as sum SNR of the components, ${\rm SNR}^2{\equiv}{\rm SNR}^2_1+{\rm SNR}^2_2$, where 
\be
{\rm SNR}^{-2}_1={\rm SNR}^{-2}_2=\sigma_\mu^2/\mu^2+\sigma_{v_{\bb,\perp}}^2/v_{\bb,\perp}^2\,,
\ee
and $\sigma_\mu^2/\mu^2=2(\sigma_{I_{\rm patch}}^2/I_{\rm patch}^2+\sigma_A^2/A^2)$ due to the angular dependence of the signal template.

\section{Angular Reconstruction}\label{sec:appendix_1}

The numerator of the signal $\tan\varphi$ can be written as 
\be\label{eq:signal__2}
\begin{split}
\alpha&=\!\!\!\!\!\!\int\limits_{\rm template}\!\!\!\!\!\dd^2\vec{r}\sin\varphi\!\!\sum\limits_{i,\rm halos}\!\!{\Psi_{u,i}}(r)\Theta^{\rm ml}(\vec{r})\\
&=\!\!\!\!\!\!\!\int\limits_{\rm template}\!\!\!\!\!\dd^2\vec{r}\sin\varphi\cos(\varphi\!-\!\varphi_0)\!\!\!\!\int\limits_{\rm patch}\!\!\!\!\dd^2\vec{R}\!\!\!\int\limits_{{z}-\rm bin}\!\!\!\!\!\dd z\,\!\!\!\!\!\!\!\!\!\!\int\limits_{\rm \ \ \ \ catalog}\!\!\!\!\!\!\!\!\!\!\dd M \\ 
&\ \ \ \ \ \ \times\left[\bar{N}_c(m_*,z)\, \chi^2\frac{\dd z}{\dd \chi}n(M,z)A(M,z)\,{\Psi_{u}}( r)\mathcal{M}_u(r)\right].
\end{split}
\ee
The denominator differs by a cosine
\be\label{eq:signal__3}
\begin{split}
\beta&=\!\!\!\!\!\!\int\limits_{\rm template}\!\!\!\!\!\dd^2\vec{r}\cos\varphi\!\!\sum\limits_{i,\rm halos}\!\!{\Psi_{u,i}}(r)\Theta^{\rm ml}(\vec{r})\\
&=\!\!\!\!\!\!\!\int\limits_{\rm template}\!\!\!\!\!\dd^2\vec{r}\cos\varphi\cos(\varphi\!-\!\varphi_0)\!\!\!\!\int\limits_{\rm patch}\!\!\!\!\dd^2\vec{R}\!\!\!\int\limits_{{z}-\rm bin}\!\!\!\!\!\dd z\,\!\!\!\!\!\!\!\!\!\!\int\limits_{\rm \ \ \ \ catalog}\!\!\!\!\!\!\!\!\!\!\dd M \\ 
&\ \ \ \ \ \ \times\left[\bar{N}_c(m_*,z)\, \chi^2\frac{\dd z}{\dd \chi}n(M,z)A(M,z)\,{\Psi_{u}}( r)\mathcal{M}_u(r)\right].
\end{split}
\ee

Errors can be calculated using a relation similar to Eq.~\eqref{eq:signal__}, with the CMB component without the moving lens effect instead, and performing the average over the realisations of the CMB as,
\be\label{eq:noise_2}
\begin{split}
\sigma_\alpha^2=\!\!\!\!\!\!\!\!&\int\limits_{\rm template}\!\!\!\!\!\!\!\dd^2\vec{r} \sin\varphi\,\dd^2\vec{r}\,'\sin\varphi'\,\\
&\times\!\!\!\!\sum_{i,j,\rm halos}\!\!\!\!\Psi_{i,u}(r)\Psi_{j,u}^*(r')\langle\tilde{\Theta}(\vec{R}_{i}\!+\!\vec{r})\tilde{\Theta}^*(\vec{R}_{j}\!+\!\vec{r}\,')\rangle\\
=\!\!\!\!\!&\!\!\!\int\limits_{\rm template}\!\!\!\!\!\!\!\dd^2\vec{r} \sin\varphi\,\dd^2\vec{r}\,'\sin\varphi'\!\!\!\!\sum_{i,j,\rm halos}\!\!\!\!\Psi_{i,u}(r)\Psi_{j,u}^*(r')\\
&\times\!\!\!\left[\iint\!\frac{\dd^2\vec{L}}{(2\pi)^2}\frac{\dd^2\vec{L'}}{(2\pi)^2}C_{L}^{\tilde{\Theta}\tilde{\Theta}}\delta^2(\vec{L}+\vec{L}')e^{-i\vec{L}\cdot(\vec{R}_i+\vec{r})-i\vec{L}'\cdot(\vec{R}_j+\vec{r}')}\right]\\
=\!\!\!\!\!&\!\!\!\int\limits_{\rm template}\!\!\!\!\!\!\!\dd^2\vec{r} \sin\varphi\,\dd^2\vec{r}\,'\sin\varphi'\!\!\!\!\sum_{i,j,\rm halos}\!\!\!\!\Psi_{i,u}(r)\Psi_{j,u}^*(r')\\
&\times\!\!\!\left[\int\!\frac{\dd^2\vec{L}}{(2\pi)^2}C_{L}^{\tilde{\Theta}\tilde{\Theta}}e^{-i\vec{L}\cdot(\vec{R}_i+\vec{r}-\vec{R}_j-\vec{r}')}\right]\\
=\!\!\!\!\!&\!\!\!\int\limits_{\rm template}\!\!\!\!\!\!\!r \dd r \,r \dd{r}\,'\!\!\!\!\sum_{i,j,\rm halos}\!\!\!\!\Psi_{i,u}(r)\Psi_{j,u}^*(r')\!\!\int\!\frac{\dd^2\vec{L}}{(2\pi)^2}C_{L}^{\tilde{\Theta}\tilde{\Theta}}e^{-i\vec{L}\cdot(\vec{R}_i-\vec{R}_j)}\\
&\times\!\!\!\left[\iint\dd\varphi\dd\varphi'e^{-irL\cos(\varphi\!-\!\varphi_0)+ir'L'\cos(\varphi'\!-\!\varphi_0)}\right]\\
=\!\!\!\!\!&\!\!\!\int\limits_{\rm template}\!\!\!\!\!\!\!r \dd r \,r \dd{r}\,'\!\!\!\!\sum_{i,j,\rm halos}\!\!\!\!\Psi_{i,u}(r)\Psi_{j,u}^*(r')\!\!\int\!\frac{\dd^2\vec{L}}{(2\pi)^2}C_{L}^{\tilde{\Theta}\tilde{\Theta}}e^{-i\vec{L}\cdot(\vec{R}_i-\vec{R}_j)}\\
&\times\!\!\left[(2\pi\sin{\varphi_0})^2\,J_1(L r)J_1(L r')\right]\,,
\end{split}\non
\ee
and the error on $\beta$ differs by a cosine and take the form, 
\be\label{eq:noise_3}
\begin{split}
\sigma_\alpha^2=\!\!\!\!\!\!\!\!&\int\limits_{\rm template}\!\!\!\!\!\!\!\dd^2\vec{r} \cos\varphi\,\dd^2\vec{r}\,'\cos\varphi'\,\\
&\times\!\!\!\!\sum_{i,j,\rm halos}\!\!\!\!\Psi_{i,u}(r)\Psi_{j,u}^*(r')\langle\tilde{\Theta}(\vec{R}_{i}\!+\!\vec{r})\tilde{\Theta}^*(\vec{R}_{j}\!+\!\vec{r}\,')\rangle\\
=\!\!\!\!\!&\!\!\!\int\limits_{\rm template}\!\!\!\!\!\!\!r \dd r \,r \dd{r}\,'\!\!\!\!\sum_{i,j,\rm halos}\!\!\!\!\Psi_{i,u}(r)\Psi_{j,u}^*(r')\!\!\int\!\frac{\dd^2\vec{L}}{(2\pi)^2}C_{L}^{\tilde{\Theta}\tilde{\Theta}}e^{-i\vec{L}\cdot(\vec{R}_i-\vec{R}_j)}\\
&\times\!\!\left[(2\pi\cos{\varphi_0})^2\,J_1(L r)J_1(L r')\right]\,,
\end{split}\non
\ee
where we defined $L=|\vec{L}|$ where $\vec{L}$ is conjugate to the radial displacement on the patch. We write the sum over the halos as, 
\be
\begin{split}
\sum_{i,j,\rm halos}\!\!\!\!&\Psi_{i,u}(r)\Psi_{j,u}^*(r')\\
=&\iint\limits_{z-\rm bin}\chi^2\chi'^2\dd\chi\,\dd\chi'\iint\limits_{\rm catalog}\dd M\dd M' n(M,\chi)n(M,\chi') \\ 
&\times\left[\Psi_{u}(r,M,\chi)\Psi_{u}^*(r',M',\chi')\iint\limits_{\rm patch}\dd^{2}\vec{R}_i\dd^{2}\vec{R}_j\right]\,,
\end{split}
\ee
and use the equality,
\be
\begin{split}
\iint\limits_{\rm patch}\dd^{2}&\vec{R}_i\dd^{2}\vec{R}_je^{-i\vec{L}\cdot(\vec{R}_i-\vec{R}_j)}\\
&=4\pi^2\!\!\iint\limits_{\rm patch}R_i R_j\dd R_i \dd R_j J_0(L R_i) J_0(L R_j)\\
&=\frac{4\pi^2}{L^2}\left[R_{\rm max} J_1(R_{\rm max} L )-R_{\rm min} J_1(R_{\rm min} L )\right]\,,
\end{split}\non
\ee
in Eqs.\eqref{eq:noise_2} and \eqref{eq:noise_3} to get Eqs.~(32-34).

\newpage

\end{document}